\documentclass[aps,pre,groupedaddress]{revtex4}
\usepackage{graphicx}
\usepackage{amsmath}
\begin{document}
\newcommand{\K}{{\mathrm K}}
\newcommand{\ud}{{\mathrm d}}
\newcommand{\umod}{\mathrm{mod}}
\newcommand{\sech}{\mathrm{sech}}
%\preprint{}

\title{Theory of frequency and phase synchronization in a rocked bistable
stochastic system }

\author{Jes\'us Casado-Pascual}
\email[]{jcasado@us.es}
\homepage[]{http://numerix.us.es}
\author{Jos\'e  G\'omez-Ord\'o\~nez}
\author{ Manuel Morillo}
\affiliation{F\'{\i}sica Te\'orica,
Universidad de Sevilla, Apartado de Correos 1065, Sevilla 41080, Spain}

%\thanks{}

\author{J\"org Lehmann}
\altaffiliation[Present address: ]{Department
f\"ur Physik und Astronomie,
Universit\"at Basel,
Klingelbergstra\ss e~82,
CH-4056~Basel, Switzerland}

\author{Igor Goychuk}
\author{Peter H\"anggi}
\affiliation{Institut f\"ur Physik,
Universit\"at Augsburg, Universit\"atsstra\ss e 1, D-86135 Augsburg,
Germany}

\date{\today}

\begin{abstract}
  We investigate the role of noise in the phenomenon of stochastic synchronization of
  switching events in a rocked, overdamped bistable potential driven
  by white Gaussian noise, the archetype description of Stochastic
  Resonance. We present a new approach to the stochastic counting
  process of noise-induced switching events: starting from the
  Markovian dynamics of the nonstationary, continuous particle
  dynamics one finds upon contraction onto two states a non-Markovian
  renewal dynamics.  The output frequency is determined as the
  velocity of the underlying discrete phase dynamics. The phenomenon
  of noise-assisted phase synchronization is investigated in terms of
  an effective, instantaneous phase diffusion. The theory is applied
  to rectangular-shaped rocking signals versus increasing input-noise
  strengths. Precise numerical simulations corroborate very favorably
  our analytical results.  The novel theoretical findings are also
  compared with prior findings.
\end{abstract}

\pacs{05.40.-a,05.45.Xt, 05.10.Gg, 02.50.-r}
%\keywords{}

\maketitle
\section{Introduction}
The theme of synchronization has wide-spread applications, covering a plenitude of
phenomena \cite{pikovskybook2001, chaos2003, vainstein1983, schimansky2001}. Some
characteristic examples are the entrainment
of  a system by external, time-dependent forcing,
or the generalization of the synchronization concept
for systems that exhibit a chaotic dynamics \cite{chaos}, lag synchronization
\cite{lag} and also phase synchronization \cite{phase}. Synchronization
phenomena play not only
a key role for diverse  technological applications but  increasingly as well
for the description, the control, and even for the therapy of selected  medical disorders
\cite{medappl}.

Due to the interaction with a surrounding environment or with internal
degrees of freedom, noise is present in many physical systems.  This
being so, its role cannot be ignored when investigating synchronization
phenomena. In recent years, it has turned out that noise can actually
play a constructive role in many physical situations. In particular,
noise can boost the transduction of information by means of the
phenomenon of Stochastic Resonance \cite{gammaitoni1998} in an ample
number of metastable physical and biological systems \cite{hanggi2002}.
Furthermore noise enables Brownian motors to do work against external
load forces \cite{BM}, or to induce phase transitions far away from
thermal equilibrium \cite{sanchobook1999,schimanskybook2002}.

Our focus here is on the role of phase synchronization in stochastic
overdamped systems driven by white Gaussian noise. In these cases, the
velocity of the dynamics is not a measurable quantity because the
stochastic trajectories are neither differentiable nor of finite
variation, see e.g.\ Ref.~\cite{HT1982}. A recently proposed method for
measuring the average phase velocity or frequency which is based on the
generalization of a Rice rate formula for threshold crossings is
consequently not a suitable method~\cite{callenbach2002}.  An
alternative approach is based on the so-called `` Hilbert phase''
dynamics, as pioneered by Gabor \cite{gabor1946} for {\it deterministic}
systems. In the present work, we shall take a closer look at the
synchronization phenomenon in a periodically driven bistable
system. Then it is advantageous to introduce a discrete phase dynamics,
as recently proposed by Schimansky-Geier and collaborators
\cite{schimansky2001, schimanskybook2002, Freund, related-items}. In
order to extract this discrete phase dynamics from the underlying
continuous process, we shall consider the stochastic counting process of the
noise-induced switches between the two potential minima. It turns out
that this counting process is in fact a non-stationary renewal
process~\cite{Cox}.

The outline of the present work is as follows: First, we introduce a
dichotomic process by filtering out the fluctuations around the
potential minima of the original stochastic process, which possesses a
clear interpretation in terms of a discrete phase. By contrast to the
underlying stochastic process, this two-state process, however, is no
longer Markovian.  Subsequently, in Sect.~III, we analyze in detail the
statistical properties of the random switching-times associated to the
dichotomic process. The one-time statistical properties of the discrete
phase are then studied in Sect.~IV. Based on these results, exact
analytical expression for the instantaneous output frequency and the
phase diffusion are derived. Approximate expressions are then derived
for the weak-noise limit and for a slow external driving. Finally, our
analytical findings will be applied to the case of a symmetric bistable
potential driven by a periodic rectangular input signal. To corroborate
our analytical results, we compare them with those obtained from a
numerical simulation of the original stochastic process.

\section{Description of the model and definition of the discrete phase}

To start, we consider a stochastic dynamics characterized by a single
degree of freedom $x(t)$, whose dynamics (in dimensionless units) is
described by the stochastic differential equation
\begin{equation}
\label{SDE}
\dot{x}(t)=-U^{\prime}\left[x(t),t\right]+\xi(t),
\end{equation}
where $\xi(t)$ is a Gaussian white noise of zero mean with
autocorrelation $\langle \xi(t) \xi(s) \rangle= 2 D \delta(t-s)$, and
$U^{\prime}(x,t)$ is the derivative with respect to $x$ of the
bistable quartic potential
\begin{equation}
\label{potential}
U(x,t)=\frac{x^4}{4}-\frac{x^2}{2}-F(t) x,
\end{equation}
$F(t)$ representing a periodic forcing with period $T$. Our focus is on
subthreshold signals; more precisely, we will assume that, for any
instant of time, $|F(t)|<A_\mathrm{th}=2/\sqrt{27}$ where
$A_\mathrm{th}$ is the static threshold value (the dynamical threshold
value always exceeds this adiabatic threshold $A_\mathrm{th}$). In this
case, the potential possesses two minima at $q_{-1}(t)<0$ and
$q_{+1}(t)>0$, and a maximum at $q_M(t)$. Introducing the function
%
%here , I changed the notation from \chi to \nu -- this dynamics
%should not be confused with your chi later, introduced in (6)!!
%
%
$\nu(t)=\arccos\left[F(t)/A_\mathrm{th}\right]$, with $\arccos y$ being the
principal value of the arc cosine of $y$ (i.e., the value in the
interval $[0,\pi]$), and
\begin{equation}
\label{locations}
\eta_n(t)= \frac{2}{\sqrt{3}} \cos\left[\frac{ \nu(t)+2 \pi n}{3}
\right],
\end{equation}
then $\eta_0(t)$ yields the location of the minimum to the right of the
barrier [i.e., $q_{+1}(t)=\eta_0(t)$], $\eta_1(t)$ yields the location
of the minimum to the left of the barrier [i.e., $q_{-1}(t)=\eta_1(t)$],
and $\eta_2(t)$ yields the location of the maximum [i.e., $q_M(t)
=\eta_2 (t)$]. From now on, we will assume that at an initial instant of
time $t_0$ the system is placed at one of the minima of the potential
$q_{\alpha_0}(t_0)$, with $\alpha_0=+1$ or $-1$. The long-time behavior
of the quantities of interest can be obtained by taking the limit as
$t_0\rightarrow -\infty$ at the end of the calculations. Henceforth, we
will make explicit the dependence of all the quantities on the initial
preparation by the superscript $\alpha_0,t_0$, see also discussion after
Eq.~\eqref{masterequation1}. Thus, for instance, we
will write $x^{\alpha_0,t_0}(t)$ instead of $x(t)$, meaning that
$x^{\alpha_0,t_0}(t_0)=q_{\alpha_0}(t_0)$.

To analyze the synchronization phenomenon in this stochastic bistable
system, it is convenient to introduce a discrete phase associated to the
continuous stochastic process $x^{\alpha_0,t_0}(t)$. In order to do so,
firstly we will proceed to filter out the fluctuations around the minima
of the stochastic process $x^{\alpha_0,t_0}(t)$ to obtain a two-state
stochastic process $\chi^{\alpha_0,t_0}(t)$ which only takes the values
$+1$ or $-1$. The procedure used is as follows: At the initial instant
of time $t_0$ we set $\chi^{\alpha_0,t_0}(t_0)=\alpha_0$. A switch of
state from $\pm \alpha_0$ to $\mp \alpha_0$ occurs whenever the system,
having started in one of the minima, reaches the other minimum for the
first time. The instant of time at which the $n$-th switch of state
takes place is a random variable which will be denoted by
$\mathcal{T}_{n}^{\alpha_0,t_0}$, with $n=1,2\dots$. Formally, these
random variables can be defined recursively as
\begin{equation}
\label{def1}
\mathcal{T}_{n}^{\alpha_0,t_0}=\min\left[t :
  t>\mathcal{T}_{n-1}^{\alpha_0,t_0} \mbox{ and }
  x^{\alpha_0,t_0}(t)=q_{\alpha_n}(t) \right],
\end{equation}
where $\mathcal{T}_{0}^{\alpha_0,t_0}=t_0$ and $\alpha_n=(-1)^n
\alpha_0$.  Thus, if we introduce the stochastic process
\begin{equation}
\label{def2}
N^{\alpha_0,t_0}(t)=\max \left[n: \mathcal{T}_{n}^{\alpha_0,t_0}\leq t
\right],
\end{equation}
which counts the number of switches of state in the interval $(t_0,t]$,
then the two-state stochastic process $\chi^{\alpha_0,t_0}(t)$ can be
expressed as
\begin{equation}
\label{defalpha}
\chi^{\alpha_0,t_0}(t)=\alpha_0 \cos\left[\pi N^{\alpha_0,t_0}(t)  \right].
\end{equation}
By analogy with the case of a sinusoidal signal, we will define the
discrete phase $\varphi^{\alpha_0, t_0}(t)$ associated with
$x^{\alpha_0,t_0}(t)$ as the stochastic process
\begin{equation}
\label{defphase}
\varphi^{\alpha_0, t_0}(t)=\pi N^{\alpha_0,t_0}(t).
\end{equation}

\begin{figure}
\includegraphics[width=8.5cm]{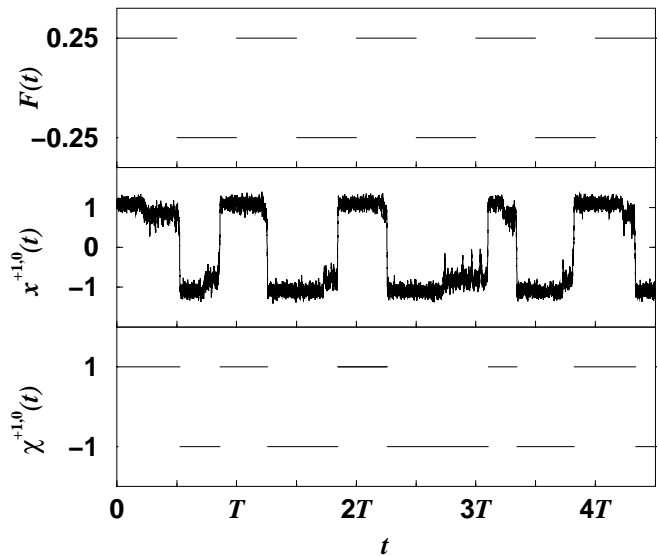}
\caption{\label{Fig1} Illustration of the procedure used to define the
  stochastic process $\chi^{\alpha_0,t_0}(t)$ in the particular case of
  a rectangular signal [see Eq.~(\ref{RF}) and text below] with
  amplitude $A=0.25$ and frequency $\Omega=2 \pi/T=0.01$, and a noise
  strength $D=0.02$. The initial instant of time has been chosen to be
  $t_0=0$, and the system has been initially placed at $q_{+1}(0)$, so
  that $\alpha_0=+1$. In the upper panel, we have sketched the
  rectangular periodic signal $F(t)$, whereas in the middle and lower
  panels, we have depicted a random trajectory of the stochastic process
  $x^{+1,0}(t)$ and the corresponding realization of the process
  $\chi^{+1,0}(t)$, respectively.}
\end{figure}

In Fig.~\ref{Fig1}, we illustrate the procedure just described in the
particular case of a rectangular signal [see Eq.~(\ref{RF}) and text
below] with amplitude $A=0.25$ and frequency $\omega=2 \pi/T=0.01$, and
a noise strength $D=0.02$. The initial instant of time has been chosen
to be $t_0=0$, and the system has been initially placed at $q_{+1}(0)$,
so that $\alpha_0=+1$. In the upper panel, we have sketched the
rectangular periodic signal $F(t)$, whereas in the middle and lower
panels, we have depicted a random trajectory of the stochastic process
$x^{+1,0}(t)$ and the corresponding realization of the process
$\chi^{+1,0}(t)$, respectively.

\section{Statistical characterization of the switching
times $\mathcal{T}_{n}^{\alpha_0,t_0}$}

According to Eqs.~(\ref{def2}) and (\ref{defphase}), the statistical
properties of the discrete phase $\varphi^{\alpha_0,t_0}(t)$ are closely
related to those of the random switching times
$\mathcal{T}_{n}^{\alpha_0,t_0}$. The aim of this section is to provide
a detailed description of the statistical characterization of these
random variables previous to the analysis of
$\varphi^{\alpha_0,t_0}(t)$, which will be postponed to the next
section. The connection between the statistical properties of the
switching times and the original stochastic process
$x^{\alpha_0,t_0}(t)$ will be also analyzed in this section.

Following the approach presented in Ref.~\cite{Cox}, the random variable
$\mathcal{T}_{n}^{\alpha_0,t_0}$ can be characterized statistically by
its probability {\it density} function
\begin{equation}
\label{defgn}
g_n^{\alpha_0,t_0}(t)=\lim_{\Delta t\rightarrow
0^+}\frac{\mathrm{Prob}\left[t<\mathcal{T}_n^{\alpha_0,t_0}\leq
t+\Delta t\right]}{\Delta t}.
\end{equation}
Besides this probability distribution function, it is also convenient to
introduce the (cumulative) distribution function
\begin{equation}
\label{defGn}
G_n^{\alpha_0,t_0}(t)=\mathrm{Prob}\left[\mathcal{T}_n^{\alpha_0,t_0}\leq
t\right]=\int_{t_0}^{t} dt^{\prime} \; g_n^{\alpha_0,t_0}(t^{\prime}),
\end{equation}
as well as its complementary,
\begin{equation}
\label{defcalGn}
\mathcal{G}_n^{\alpha_0,t_0}(t)=\mathrm{Prob}\left[\mathcal{T}_n^{\alpha_0,t_0}>
t\right]=1-G_n^{\alpha_0,t_0}(t).
\end{equation}

In the particular case $n=1$, these functions can be directly determined
from the solution of the Fokker-Planck equation (FPE)
\begin{equation}
\label{FPE}
\frac{\partial}{\partial t} P\left(x,t\right) =\frac{\partial}{\partial
x}\left[D\frac{\partial}{\partial x}+U^{\prime}(x,t)
\right]P\left(x,t\right),
\end{equation}
with initial condition
\begin{equation}
\label{ic}
P\left(x,t_0\right)=\delta\left[x-q_{\alpha_0}(t_0) \right],
\end{equation}
and absorbing boundary condition at $q_{-\alpha_0}(t)$, i.e.,
\begin{equation}
\label{abc}
P\left[q_{-\alpha_{0}}(t),t\right]=0 \text{ for all $t\ge t_0$.}
\end{equation}
Denoting by $P^{\alpha_0,t_0}\left(x,t\right)$ the solution of the above
problem, it follows from the definition of
$\mathcal{G}_1^{\alpha_0,t_0}(t)$ in Eq.~(\ref{defcalGn}), with $n=1$,
that
\begin{equation}\label{add}
\mathcal{G}_1^{\alpha_0,t_0}(t)= \delta_{\alpha_0,+1}
\int_{q_{-1}(t)}^{+\infty} d x \,P^{+1,t_0}\left(x,t\right)
+\delta_{\alpha_0,-1} \int_{-\infty}^{q_{+1}(t)} d x
\,P^{-1,t_0}\left(x,t\right),
\end{equation}
for $t\geq t_0$, and $\mathcal{G}_1^{\alpha_0,t_0}(t)=1$ for
$t < t_0$.
%%%%%%%%
The function $\mathcal{G}_1^{\alpha_0,t_0}(t)$ is the
conditional survival
probability of the discrete state $\alpha_0$.
The function $G_1^{\alpha_0,t_0}(t)$ is then given by
Eq.~(\ref{defcalGn}) with $n=1$, whereas
$g_1^{\alpha_0,t_0}(t)=-\dot{\mathcal{G}}_1^{\alpha_0,t_0}(t)$
is the corresponding conditional residence time distribution (RTD). The dot
indicates the derivative with respect to time $t$.
The knowledge of either the conditional survival probabilities
$\mathcal {G}_1^{\alpha_0,t_0}(t)$, or (equivalently)
the conditional RTDs $g_1^{\alpha_0,t_0}(t)$
is sufficient to specify a driven two state non-Markovian
renewal process  \cite{GH04}.
These functions can be found from the underlying continuous state
Markovian dynamics by solving Eqs. (\ref{FPE})-(\ref{add}).
%%%%%%%

For $n>1$, the functions $\mathcal{G}_n^{\alpha_0,t_0}(t)$,
$G_n^{\alpha_0,t_0}(t)$, and $g_n^{\alpha_0,t_0}(t)$ can be obtained
iteratively from the ones corresponding to the case $n=1$ by making use
of three integral equations. To obtain the first integral equation, let
us consider the consistency condition
\begin{equation}
\label{p1}
\mathcal{G}_{n+1}^{\alpha_0,t_0}(t)=\int_{-\infty}^{+\infty} d
t^{\prime} \;\mathrm{Prob}\left[\mathcal{T}_{n+1}^{\alpha_0,t_0}>
t\left| \right. \mathcal{T}_{n}^{\alpha_0,t_0}=t^{\prime} \right]
g_n^{\alpha_0,t_0}(t').
\end{equation}
By use of the definition of the switching times in Eq.~(\ref{def1}), as well
as the Markovian character of the original stochastic process
$x^{\alpha_0,t_0}(t)$, it is straightforward to verify that
\begin{equation}
\mathrm{Prob}\left[\mathcal{T}_{n+1}^{\alpha_0,t_0}> t\left|
\right. \mathcal{T}_{n}^{\alpha_0,t_0}=t^{\prime}
\right]=\mathcal{G}_1^{\alpha_n,t^{\prime}}(t).
\end{equation}
Inserting the above expression into Eq.~(\ref{p1}) and taking into
account that $\mathcal{G}_1^{\alpha_n,t^{\prime}}(t)=1$ for
$t<t^{\prime}$ and $g_n^{\alpha_0,t_0}(t)=0$ for $t<t_0$, we obtain
\begin{equation}
\label{IntEqcalGn}
\mathcal{G}_{n+1}^{\alpha_0,t_0}(t)=\mathcal{G}_{n}^{\alpha_0,t_0}(t)+\int_{t_0}^{t}
d t^{\prime}\;
\mathcal{G}_1^{\alpha_n,t^{\prime}}(t)g_{n}^{\alpha_0,t_0}(t^{\prime}),
\end{equation}
%
%here...your first formulation was cryptic to me ...a rephrased ...please check....
%
for $n\geq 1$. The interpretation of this result is straightforward: The
probability that the $(n+1)$-th switch of state occurs after the
time~$t$ is equal to the probability that the $n$-th switch occurs after
that instant of time plus the probability that the $n$-th switch
has happened at any time $t'$ before $t$ with the next switch taking place
after $t$. Similar interpretations hold for the integral equations;
i.e.,
\begin{equation}
\label{IntEqGn}
G_{n+1}^{\alpha_0,t_0}(t)= \int_{t_0}^{t} dt^{\prime}\;G_1^{\alpha_n,t^{\prime}}(t)
g_n^{\alpha_0,t_0}(t'),
\end{equation}
\begin{equation}
\label{IntEqgn}
g_{n+1}^{\alpha_0,t_0}(t)= \int_{t_0}^{t} dt^{\prime}\;g_1^{\alpha_n,t^{\prime}}(t)
g_n^{\alpha_0,t_0}(t'),
\end{equation}
which are obtained from Eq.~(\ref{IntEqcalGn}) by using
Eq.~(\ref{defcalGn}) and
$g_n^{\alpha_0,t_0}(t)=-\dot{\mathcal{G}}_n^{\alpha_0,t_0}(t)$,
respectively.

A formal solution for $g_{n}^{\alpha_0,t_0}(t_{n})$ is obtained by
solving iteratively the integral equation~(\ref{IntEqgn}). The result is
\begin{equation}
\label{res1}
g_{n}^{\alpha_0,t_0}(t_{n})=\int_{t_0}^{t_{n}}dt_{n-1}\dots\int_{t_0}^{t_2}
dt_{1}\; \prod_{j=0}^{n-1} g_1^{\alpha_j,t_j}(t_{j+1}),
\end{equation}
for $n\geq 2$. Thus, the probability distribution corresponding to the
first switch of state, $g_1^{\alpha_0,t_0}(t)$, which can be obtained from
the solution of the FPE (\ref{FPE}) with (\ref{ic}) and (\ref{abc}),
determines completely the statistical properties of the rest of the
switching times.

\section{One-time statistical properties of the discrete phase: the
  output frequency and the phase dispersion}

The one-time statistical properties of the discrete phase
$\varphi^{\alpha_0,t_0}(t)$ can be evaluated by making use of the
probability distribution of the number of switches of state
\begin{equation}
\label{defpn}
\rho_n^{\alpha_0,t_0}(t)=\mathrm{Prob}\left[N^{\alpha_0,t_0}(t)=n
\right],
\end{equation}
with $n=0,1,2,\dots$. From the definition of $N^{\alpha_0,t_0}(t)$ in
Eq.~(\ref{def2}) it follows that
\begin{equation}
\label{prop1}
\mathrm{Prob}\left[N^{\alpha_0,t_0}(t)\geq n
\right]=G_n^{\alpha_0,t_0}(t),
\end{equation}
with $G_0^{\alpha_0,t_0}(t)=1$. Consequently, the probability
distribution of the number of switches of state and its derivative with
respect to $t$ can be expressed, respectively, as
\begin{eqnarray}
\label{pn1}
\rho_n^{\alpha_0,t_0}(t)&=&G_n^{\alpha_0,t_0}(t)-G_{n+1}^{\alpha_0,t_0}(t)
\nonumber\\
&=&\mathcal{G}_{n+1}^{\alpha_0,t_0}(t)-\mathcal{G}_{n}^{\alpha_0,t_0}(t),
\end{eqnarray}
and
\begin{equation}
\label{meq1}
\dot{\rho}_n^{\alpha_0,t_0}(t)=g_n^{\alpha_0,t_0}(t)-g_{n+1}^{\alpha_0,t_0}(t),
\end{equation}
with $\mathcal{G}_0^{\alpha_0,t_0}(t)=g_0^{\alpha_0,t_0}(t)=0$. The
average of an arbitrary one-time function of $N^{\alpha_0,t_0}(t)$,
$K\left[N^{\alpha_0,t_0}(t)\right]$, is obviously given by
\begin{equation}
\label{average}
\left\langle K\left[N^{\alpha_0,t_0}(t) \right]\right\rangle =\sum_{n=0}^{\infty}
K(n)\rho_n^{\alpha_0,t_0}(t).
\end{equation}

Equations~(\ref{pn1}) and (\ref{meq1}) can be written in a more
transparent form by introducing the probability of an almost immediate
switch of state after $n$ switches
\begin{equation}
\label{defrate1}
\Gamma_{n}^{\alpha_0,t_0}(t)=\lim_{\Delta t\rightarrow 0^{+}}
\frac{\mathrm{Prob}\left[t<\mathcal{T}_{n+1}^{\alpha_0,t_0}
\leq t+\Delta t\left| \right.N^{\alpha_0,t_0}(t)=n
\right]}{\Delta t}.
\end{equation}
Note that if the process $\chi^{\alpha_0,t_0}(t)$ were Markovian, these
probabilities could only depend on the state $\alpha_n$. The explicit
dependence on the number of jumps $n$ and on the initial preparation
$\alpha_0$ at time $t_0$ is a consequence of the non-Markovian character
of the process. Another fingerprint of the non-Markovian nature of the
dichotomic process is the fact that these probabilities depend on the
time $t$ even in the absence of the external driving. In order to
clarify this point, let us consider, e.g., the particular case
$n=0$. Then, while initially, right after the particle has been prepared
at one of the minima, the distribution function $P^{\alpha_0,t_0}(x, t)$
is still very sharply peaked around $q_{\alpha_0}(t_0)$, it becomes
smeared out around the minimum after the intra-well relaxation time.
Consequently, the probability of an immediate switch is obviously will
be different in both situations, even without an external driving.

Multiplying and dividing the right-hand side of the above expression by
$\rho_n^{\alpha_0,t_0}(t)$ and taking into account that
\begin{equation}
\lim_{\Delta t\rightarrow 0^{+}} \frac{\mathrm{Prob}\left[t<\mathcal{T}_{n+1}^{\alpha_0,t_0} \leq t+\Delta
t\mbox{ and } N^{\alpha_0,t_0}(t)=n
\right]}{\Delta t}=g_{n+1}^{\alpha_0,t_0}(t)
\end{equation}
it is readily seen that
\begin{equation}
\label{defrate2}
\Gamma_{n}^{\alpha_0,t_0}(t)=\frac{g_{n+1}^{\alpha_0,t_0}(t)}{\rho_n^{\alpha_0,t_0}(t)}\,.
\end{equation}
Then, it follows from Eqs.~(\ref{IntEqcalGn}), (\ref{pn1}), and
\eqref{defrate2} that
\begin{equation}
\label{pn2}
\rho_n^{\alpha_0,t_0}(t)=\int_{t_0}^{t} d t^{\prime}\;
\mathcal{G}_1^{\alpha_n,t^{\prime}}(t)\Gamma_{n-1}^{\alpha_0,t_0}(t^{\prime})\rho_{n-1}^{\alpha_0,t_0}(t^{\prime}),
\end{equation}
for $n\geq 1$, with
$\rho_0^{\alpha_0,t_0}(t)=\mathcal{G}_1^{\alpha_0,t_0}(t)$. Analogously,
Eq.~(\ref{meq1}) leads to the following hierarchy of differential
equations
\begin{equation}
\label{meq2}
\dot{\rho}_n^{\alpha_0,t_0}(t)=\Gamma_{n-1}^{\alpha_0,t_0}(t)\rho_{n-1}^{\alpha_0,t_0}(t)-\Gamma_{n}^{\alpha_0,t_0}(t)\rho_{n}^{\alpha_0,t_0}(t),
\end{equation}
for $n\geq 1$, and
\begin{equation}
\label{meq3}
\dot{\rho}_0^{\alpha_0,t_0}(t)=-\Gamma_{0}^{\alpha_0,t_0}(t)\rho_{0}^{\alpha_0,t_0}(t),
\end{equation}
which must be solved with the initial condition
$\rho_{n}^{\alpha_0,t_0}(t_0)=\delta_{n,0}$.

%%%%%
We will also introduce the conditional probability for
$\chi^{\alpha_0,t_0}(t)$ to take the value $\beta=\pm 1$ at time
$t$, provided that it took the value $\alpha_0$ with probability one
at the initial instant of time $t_0$,
\begin{equation}
\label{defp}
p_{\beta}^{\alpha_0,t_0}(t)=\mathrm{Prob}\left[\chi^{\alpha_0,t_0}(t)=\beta\right].
\end{equation}
%%%%%%%
Noting that after an even number of
switches of state the system ends up in the same state as it was
initially, whereas for an odd number of switches the system ends up in
the other state, it is clear that the events
$\left\{\chi^{\alpha_0,t_0}(t)=\alpha_0\right\}$ and
$\left\{\chi^{\alpha_0,t_0}(t)=-\alpha_0\right\}$ are respectively
equivalent to the events $\left\{N^{\alpha_0,t_0}(t)\mbox{ is
even}\right\}$ and $\left\{N^{\alpha_0,t_0}(t) \mbox{ is odd}\right\}$,
and consequently
\begin{equation}
\label{defpbeta}
p_{\beta}^{\alpha_0,t_0}(t)= \delta_{\alpha_0,\beta}\sum_{n=0}^{\infty}\rho_{2
n}^{\alpha_0,t_0}(t)+\delta_{-\alpha_0,\beta} \sum_{n=0}^{\infty}\rho_{2
n+1}^{\alpha_0,t_0}(t).
\end{equation}

Besides the probability distribution of the number of switches of state,
later we will also use the probability distribution of the number of
switches of state conditioned to the value of $\chi^{\alpha_0,t_0}(t)$,
\begin{equation}
\label{condpndef}
\rho_{n}^{\alpha_0,t_0}(t|\beta)=\mathrm{Prob}\left[N^{\alpha_0,t_0}(t)=n\left|
  \right. \chi^{\alpha_0,t_0}(t)=\beta \right].
\end{equation}
Multiplying and dividing the right-hand side of the above expression by
$p_{\beta}^{\alpha_0,t_0}(t)$ and taking into account that
$\mathrm{Prob}\left[N^{\alpha_0,t_0}(t)=n \mbox{ and }
\chi^{\alpha_0,t_0}(t)=\beta
\right]=\rho_{n}^{\alpha_0,t_0}(t)\delta_{\alpha_n,\beta}$, it results
that
\begin{equation}
\label{condpn}
\rho_{n}^{\alpha_0,t_0}(t|\beta)=
\frac{\rho_{n}^{\alpha_0,t_0}(t)}{p_{\beta}^{\alpha_0,t_0}(t)}\;\delta_{\alpha_n,\beta}.
\end{equation}
The average of an arbitrary one-time function of $N^{\alpha_0,t_0}(t)$,
$K\left[N^{\alpha_0,t_0}(t)\right]$, conditioned to the event
$\left\{\chi^{\alpha_0,t_0}(t)=\beta\right\}$ will be denoted by
$\left\langle K\left[N^{\alpha_0,t_0}(t)\right] \right\rangle_{\beta}$,
with $\beta=+1 \mbox{ or } -1$, and it is given by
\begin{eqnarray}
\label{condaverage}
\left\langle K\left[N^{\alpha_0,t_0}(t)\right] \right\rangle_{\beta}&=&
\sum_{n=0}^{\infty} K(n) \rho_{n}^{\alpha_0,t_0}(t|\beta)\nonumber
\\&=&\delta_{\alpha_0,\beta}\sum_{n=0}^{\infty}K(2n)\frac{\rho_{2
n}^{\alpha_0,t_0}(t)}{p_{\alpha_0}^{\alpha_0,t_0}(t)}+\delta_{-\alpha_0,\beta}
\sum_{n=0}^{\infty}K(2n+1)\frac{\rho_{2
n+1}^{\alpha_0,t_0}(t)}{p_{-\alpha_0}^{\alpha_0,t_0}(t)}.
\end{eqnarray}
>From the above expression and Eq.~(\ref{average}), it follows that
\begin{equation}
\label{average2}
\left\langle K\left[N^{\alpha_0,t_0}(t) \right]\right\rangle
=\left\langle K\left[N^{\alpha_0,t_0}(t)\right]
\right\rangle_{+1}p_{+1}^{\alpha_0,t_0}(t)+\left\langle
K\left[N^{\alpha_0,t_0}(t)\right]
\right\rangle_{-1}p_{-1}^{\alpha_0,t_0}(t).
\end{equation}

Another interesting quantity which will be useful later is the
probability of an almost immediate switch from state $\beta$, defined as
\begin{equation}
\label{defrate3}
\gamma_{\beta}^{\alpha_0,t_0}(t)=\lim_{\Delta t\rightarrow 0^{+}}
\frac{\mathrm{Prob}\left[t<\mathcal{T}_{N^{\alpha_0,t_0}(t)+1}^{\alpha_0,t_0}
\leq t+\Delta t\left| \right. \chi^{\alpha_0,t_0}(t)=\beta
\right]}{\Delta t}.
\end{equation}
Notice that in the Markovian limit these probability are independent of
the initial preparation $\alpha_0$ at time $t_0$ and cannot be
distinguished from the probabilities defined in Eq.~\eqref{defrate1}.
Multiplying and dividing the right-hand side of the above expression by
$p_{\beta}^{\alpha_0,t_0}(t)$ and taking into account the equivalence of
the event $\left\{\chi^{\alpha_0,t_0}(t)=\beta \right\}$ with the event
$\left\{N^{\alpha_0,t_0}(t)\mbox{ is
even}\right\}\delta_{\alpha_0,\beta}+\left\{N^{\alpha_0,t_0}(t)\mbox{ is
odd}\right\}\delta_{-\alpha_0,\beta}$, as well as Eqs.~(\ref{defrate1})
and (\ref{condaverage}), it is easy to see that
\begin{equation}
\label{rates1}
\gamma_{\beta}^{\alpha_0,t_0}(t)=\left\langle
\Gamma_{N^{\alpha_0,t_0}(t)}^{\alpha_0,t_0}(t) \right \rangle_{\beta}.
\end{equation}

Differentiating Eq.~(\ref{defpbeta}) with respect to $t$ and taking into
account Eqs.~(\ref{meq2}), (\ref{condaverage}), and (\ref{rates1}), it is
straightforward to obtain that
\begin{equation}
\label{masterequation1}
\dot{p}_{\beta}^{\alpha_0,t_0}(t)=-\gamma_{\beta}^{\alpha_0,t_0}(t)
p_{\beta}^{\alpha_0,t_0}(t)+\gamma_{-\beta}^{\alpha_0,t_0}(t)
p_{-\beta}^{\alpha_0,t_0}(t),
\end{equation}
for $\beta=1 \mbox{ and } -1$.
%%%%%%%%%%%%%%%%%%%%%%%%%%%%%%%%%%%%%%%
The Eq. (\ref{masterequation1}) is a non-Markovian master equation
for the
conditional probabilities $p_{\beta}^{\alpha_0,t_0}(t)$. It is of
the time-convolutionless
form \cite{Hanggi1977,Hanggi1978}. The rate
parameters $\gamma_{\beta}^{\alpha_0,t_0}(t)$ entering this
equation are time-dependent quantities even in the absence of
time-dependent driving. Such a time-dependence reflects primarily a
non-exponential distribution of the residence times of the renewal
two state non-Markovian process \cite{Cox,Kampen79} which results
from the projection of a continuous state Markovian stochastic dynamics
onto the two discrete states $\beta=\pm 1$. A time-dependent
driving introduces an additional time dependence into
$\gamma_{\beta}^{\alpha_0,t_0}(t)$ which is present also in the
{\it driven} Markovian case. In this latter case,
$\gamma_{\beta}^{\alpha_0,t_0}(t)$ becomes a time-dependent rate and
this rate depends neither on $\alpha_0$,
nor on $t_0$ (see below). Two other forms are possible to describe the
evolution of conditional probabilities $p_{\beta}^{\alpha_0,t_0}(t)$.
One is given by the generalized master equations (GMEs) with the
memory kernels
expressed  via the corresponding RTDs.
In a driven case, the kernels of corresponding GMEs will become
functionals of driving and will depend on
both time arguments.
Alternatively, the integral equations for the conditional probabilities
$p_{\beta}^{\alpha_0,t_0}(t)$ can be derived for the driven two-state
renewal process in terms of (conditional) RTDs $g_1^{\alpha_0,t_0}(t)$
\cite{GH04}. Such integral equations present
a generalization of the integral renewal equations of Ref.
\cite{Cox} onto the driven
case. We apply in this work a {\it time-convolutionless}
description  of non-Markovian dynamics \cite{Hanggi1977,Hanggi1978}
to the synchronization problem.

%%%%%%%%%%%%%%%%%%%%%%%%%%%%%%%%%%%%%%%%

%It is important to emphasize again
%that, despite the existence of a master-equation-like formulation for
%the one-time probability $ p_{\beta}^{\alpha_0,t_0}(t)$, the
%stochastic process $\chi^{\alpha_0,t_0}(t)$ is
%non-Markovian~\cite{Hanggi1977,Hanggi1978}. The
%non-Markovian character of the process $\chi^{\alpha_0,t_0}(t)$
%becomes evident if we consider the two conditional probabilities
%$p^{\alpha_0, t_0}_\beta(t|-\beta,t_1, \beta, t_2)$ and $p^{\alpha_0,
%t_0}_\beta(t|-\beta,t_1)$ for $t>t_1>t_2>t_0$, and $t-t_2$ being much
%smaller than the characteristic intra-well relaxation times.  Then,
%the former conditional probability almost vanishes, because the
%particle has just arrived at the minimum $q_{-\beta}(t_1)$ and
%consequently the probability distribution $P^{\alpha_0, t_0}(x,t)$ is
%still very peaked around the minimum. On the other hand, a similar
%argument cannot be applied to the latter conditional probability,
%which therefore is non-zero \textit{a priori}. Note that this argument
%is independent of the preparation time $t_0$ and in particular also
%holds in the limit $t_0\to-\infty$, when the system has forgotten the
%initial preparation. At this point, we also want to point out that
%there is a crucial difference between specifying the state of the
%underlying process $x^{\alpha_0, t_0}(t)$, which we have denoted by a
%superscript, and specifying only the state of the dichotomic process
%$\chi^{\alpha_0, t_0}(t)$, denoted above as a conditional probability.

After these rather formal considerations, we shall now apply these
results to the evaluation of two important quantities in the study of
the synchronization phenomenon: The instantaneous output frequency and
phase diffusion.

\subsection{The instantaneous output frequency}

The instantaneous output frequency is defined as the derivative with
respect to $t$ of the averaged discrete phase \cite{Freund}, i.e.,
\begin{equation}
\label{defoutputfreq}
\Omega_\mathrm{out}^{\alpha_0,t_0}(t)=\frac{\partial}{\partial t}\left\langle
\varphi^{\alpha_0,t_0}(t)\right\rangle=\pi \frac{\partial}{\partial t}\left\langle
N^{\alpha_0,t_0}(t)\right\rangle.
\end{equation}
Multiplying Eq.~(\ref{meq2}) by $n$, summing up the series
$\sum_{n=1}^{\infty}n \dot{\rho}_n^{\alpha_0,t_0}(t)$, and taking into
account Eq.~(\ref{average}), it is easy to obtain that
\begin{equation}
\label{eqres1}
\frac{\partial}{\partial t}\left\langle
N^{\alpha_0,t_0}(t)\right\rangle=\left\langle
\Gamma_{N^{\alpha_0,t_0}(t)}^{\alpha_0,t_0}(t)\right\rangle.
\end{equation}
Consequently, from Eqs.~(\ref{average2}), (\ref{rates1}), and
(\ref{defoutputfreq}), it results
\begin{equation}
\label{freqoutres}
\Omega_\mathrm{out}^{\alpha_0,t_0}(t)=\pi \left[\gamma_{+1}^{\alpha_0,t_0}(t)
p_{+1}^{\alpha_0,t_0}(t)+\gamma_{-1}^{\alpha_0,t_0}(t)
p_{-1}^{\alpha_0,t_0}(t)\right].
\end{equation}
This finding for the averaged frequency of the discrete phase dynamics
constitutes a first main finding  of this work.

\subsection{The instantaneous phase diffusion}

Let us now proceed to the evaluation of the instantaneous phase
diffusion $D_\mathrm{out}^{\alpha_0,t_0}(t)$, which can be defined as \cite{Freund}
\begin{eqnarray}
\label{phasedispdef1}
D_\mathrm{out}^{\alpha_0,t_0}(t)&=&\frac{\partial}{\partial t}
 \left\{\left\langle \left[
 \varphi^{\alpha_0,t_0}(t)\right]^2\right\rangle-\left\langle
 \varphi^{\alpha_0,t_0}(t)\right\rangle^2\right\}\nonumber\\ &=&\pi^2
 \frac{\partial}{\partial t} \left\{\left\langle \left[
 N^{\alpha_0,t_0}(t)\right]^2\right\rangle-\left\langle
 N^{\alpha_0,t_0}(t)\right\rangle^2\right\}.
\end{eqnarray}
Multiplying Eq.~(\ref{meq2}) by $n^2$, summing up the series
$\sum_{n=1}^{\infty}n^2 \dot{\rho}_n^{\alpha_0,t_0}(t)$, and taking into
account Eq.~(\ref{average}), it is straightforward to see that
\begin{equation}
\label{phasedis1}
\frac{\partial}{\partial t}\left\langle\left[
 N^{\alpha_0,t_0}(t)\right]^2\right\rangle=2\left\langle
 N^{\alpha_0,t_0}(t)
 \Gamma_{N^{\alpha_0,t_0}(t)}^{\alpha_0,t_0}(t)\right\rangle+\left\langle
 \Gamma_{N^{\alpha_0,t_0}(t)}^{\alpha_0,t_0}(t)\right\rangle.
\end{equation}
Replacing the above expression into Eq.~(\ref{phasedispdef1}) and taking
into account Eqs.~(\ref{defoutputfreq}) and (\ref{eqres1}), it results
that
\begin{eqnarray}
\label{phasedis2}
D_\mathrm{out}^{\alpha_0,t_0}(t)&=&\pi \,\Omega_\mathrm{out}^{\alpha_0,t_0}(t)\nonumber\\&&+2 \pi^2
\left[\left\langle N^{\alpha_0,t_0}(t)
\Gamma_{N^{\alpha_0,t_0}(t)}^{\alpha_0,t_0}(t)\right\rangle-\left\langle
N^{\alpha_0,t_0}(t)\right\rangle
\left\langle\Gamma_{N^{\alpha_0,t_0}(t)}^{\alpha_0,t_0}(t)\right\rangle\right].
\end{eqnarray}
Equation~(\ref{phasedis2}) can be expressed in a more convenient form by
writing all the averages $\langle \dots \rangle$ in term of the
conditional averages $\langle \dots \rangle_{\beta}$, according to
Eq.~(\ref{average2}). Then, after some simplifications one obtains
\begin{eqnarray}
\label{phasedis3}
D_\mathrm{out}^{\alpha_0,t_0}(t)&=&\pi\,
\Omega_\mathrm{out}^{\alpha_0,t_0}(t)+2\pi^2\Delta\gamma^{\alpha_0,t_0}(t)\Psi^{\alpha_0,t_0}(t)\nonumber
\\&&+2\pi^2\sum_{\beta=\pm
1}C_{\beta}^{\alpha_0,t_0}(t)p_{\beta}^{\alpha_0,t_0}(t),
\end{eqnarray}
where
\begin{equation}
\label{defdgamma}
\Delta\gamma^{\alpha_0,t_0}(t)=\gamma_{+1}^{\alpha_0,t_0}(t)-\gamma_{-1}^{\alpha_0,t_0}(t),
\end{equation}
\begin{equation}
\label{defPsi}
\Psi^{\alpha_0,t_0}(t)=\left[\left\langle N^{\alpha_0,t_0}(t)
\right\rangle_{+1}-\left\langle N^{\alpha_0,t_0}(t)
\right\rangle_{-1}\right]
p_{+1}^{\alpha_0,t_0}(t)p_{-1}^{\alpha_0,t_0}(t),
\end{equation}
and we have introduced the conditional covariance
\begin{equation}
\label{defC}
C_{\beta}^{\alpha_0,t_0}(t)=\left\langle
\left[N^{\alpha_0,t_0}(t)-\left\langle
N^{\alpha_0,t_0}(t)\right\rangle_{\beta}\right]
\left[\Gamma_{N^{\alpha_0,t_0}(t)}^{\alpha_0,t_0}(t)-\left\langle
\Gamma_{N^{\alpha_0,t_0}(t)}^{\alpha_0,t_0}(t)\right\rangle_{\beta}\right]\right\rangle_{\beta}.
\end{equation}

Making use of Eqs.~(\ref{meq2}) and (\ref{masterequation1}) and after
some lengthy calculations, it is possible to prove that
$\Psi^{\alpha_0,t_0}(t)$ satisfies the differential equation
\begin{eqnarray}
\label{phasedis4}
\dot{\Psi}^{\alpha_0,t_0}(t)&=&-\gamma^{\alpha_0,t_0}(t)
\Psi^{\alpha_0,t_0}(t)\nonumber\\ &&-\sum_{\beta=\pm 1}\beta
\,\left[C_{\beta}^{\alpha_0,t_0}(t)+\gamma_{\beta}^{\alpha_0,t_0}(t)
p_{\beta}^{\alpha_0,t_0}(t)\right] p_{\beta}^{\alpha_0,t_0}(t),
\end{eqnarray}
where
\begin{equation}
\label{phasedis5}
\gamma^{\alpha_0,t_0}(t)=\gamma_{+1}^{\alpha_0,t_0}(t)+\gamma_{-1}^{\alpha_0,t_0}(t).
\end{equation}
Equation~(\ref{phasedis4}) can be formally solved taking into account
that, as it follows from the definition~(\ref{defPsi}),
$\Psi^{\alpha_0,t_0}(t_0)=0$. The result is
\begin{eqnarray}
\label{phasedis6}
\Psi^{\alpha_0,t_0}(t)&=&-\sum_{\beta=\pm 1} \beta \int_{t_0}^{t} d
t^{\prime}
\,\left[C_{\beta}^{\alpha_0,t_0}(t^{\prime})+\gamma_{\beta}^{\alpha_0,t_0}(t^{\prime})
p_{\beta}^{\alpha_0,t_0}(t^{\prime})\right]\nonumber\\ && \times
p_{\beta}^{\alpha_0,t_0}(t^{\prime}) e^{-\int_{t^{\prime}}^{t} d
t^{\prime \prime} \gamma^{\alpha_0,t_0}(t^{\prime\prime})}.
\end{eqnarray}
Replacing the above expression into Eq.~(\ref{phasedis3}), one obtains
\begin{eqnarray}
\label{phasedis7}
D_\mathrm{out}^{\alpha_0,t_0}(t)&=&\pi\,
\Omega_\mathrm{out}^{\alpha_0,t_0}(t)+Q^{\alpha_0,t_0}(t)
\nonumber\\&&-2\pi^2\Delta\gamma^{\alpha_0,t_0}(t)\sum_{\beta=\pm 1}
\beta \int_{t_0}^{t} d t^{\prime}
\gamma_{\beta}^{\alpha_0,t_0}(t^{\prime})
\left[p_{\beta}^{\alpha_0,t_0}(t^{\prime})\right]^2
e^{-\int_{t^{\prime}}^{t} d t^{\prime \prime}
\gamma^{\alpha_0,t_0}(t^{\prime\prime})},
\end{eqnarray}
where all the dependence on the conditional covariance
$C_{\beta}^{\alpha_0,t_0}(t)$ has been included in the function
\begin{eqnarray}
\label{phasedis8}
Q^{\alpha_0,t_0}(t)&=&2\pi^2\sum_{\beta=\pm
1}C_{\beta}^{\alpha_0,t_0}(t)p_{\beta}^{\alpha_0,t_0}(t) \nonumber \\
&&-2\pi^2\Delta\gamma^{\alpha_0,t_0}(t)\sum_{\beta=\pm 1} \beta
\int_{t_0}^{t} d t^{\prime} C_{\beta}^{\alpha_0,t_0}(t^{\prime})
p_{\beta}^{\alpha_0,t_0}(t^{\prime}) e^{-\int_{t^{\prime}}^{t} d
t^{\prime \prime} \gamma^{\alpha_0,t_0}(t^{\prime\prime})}.
\end{eqnarray}
The expresion~(\ref{phasedis7}) for the instantaneous phase diffusion
presents a  second main result of this work.

\subsection{The weak-noise and low-frequency limit}

Throughout the following, we will assume that the noise strength~$D$
is sufficiently small so that the intrawell relaxation time scale is
negligible compared with the time scale associated to the interwell
transitions and, as well, the driving time scale $T$. In this case,
for $t-t_0$ much larger than the characteristic intrawell relaxation
time, the probability of an almost immediate switch of state after $0$
switches, $\Gamma_0^{\beta,t_0}(t)$, can be approximated by the
Kramers rate of escape \cite{Hanggi1} from the state $\beta$ at time
$t$, i.e.,
\begin{equation}
\label{Kramers}
\Gamma_0^{\beta,t_0}(t)\approx
\gamma_\beta^{\K}(t)=\frac{\omega_\beta(t) \, \omega_M(t)}{2 \pi}
\exp\left\{-\frac{U\left[q_M(t),t\right]-U\left[q_\beta(t),t\right]}{D}
\right\},
\end{equation}
where $\omega_\beta(t)=\sqrt{U''[q_\beta(t),t]}=\sqrt{3
[q_\beta(t)]^2-1}$ and $\omega_M(t)=\sqrt{|U''[q_M(t),t]|}=\sqrt{1-3
[q_M(t)]^2}$. Furthermore, from Eqs.~\eqref{IntEqcalGn},
\eqref{IntEqgn}, \eqref{pn1}, and \eqref{defrate2} it follows that,
within this approximation, we also have that
$\Gamma_n^{\beta,t_0}(t)\approx \gamma_\beta^{\K}(t)$ for $n\geq 1$ and,
consequently, $\gamma_\beta^{\alpha_0,t_0}(t)\approx
\gamma_\beta^{\K}(t)$ and $C_{\beta}^{\alpha_0,t_0}(t)\approx 0$.
%%%%%%%%%%%%%%% 2 new equations %%%%
In this case, the conditional survival propabilities and the residence time
distributions in state $\alpha_0$ read
\begin{eqnarray}
\mathcal{G}_1^{\alpha_0,t_0}(t)=\exp \Big [-\int_{t_0}^t
\gamma^K_{\alpha_0}(t')dt' \Big]
\end{eqnarray}
and
\begin{eqnarray}
g_1^{\alpha_0,t_0}(t)= \gamma^K_{\alpha_0}(t)\exp \Big [-\int_{t_0}^t
\gamma^K_{\alpha_0}(t')dt' \Big],
\end{eqnarray}
respectively. This corresponds to a two-state Markovian process
with rates $\gamma^K_{\pm 1}(t)$.
In this Markovian limit for the reduced, two-state dynamics,
the instantaneous output frequency and phase diffusion
become, % exactly, without further approximations involved
\begin{equation}
\label{freqoutres1}
\Omega_\mathrm{out}^{\alpha_0,t_0}(t)=\pi \left[\gamma_{+1}^{\K}(t)
p_{+1}^{\alpha_0,t_0}(t)+\gamma_{-1}^{\K}(t)
p_{-1}^{\alpha_0,t_0}(t)\right],
\end{equation}
and
\begin{eqnarray}
\label{phasedis9}
D_\mathrm{out}^{\alpha_0,t_0}(t)&=&\pi\,
\Omega_\mathrm{out}^{\alpha_0,t_0}(t)
\nonumber\\&&-2\pi^2\Delta\gamma^{\K}(t)\sum_{\beta=\pm 1}
\beta \int_{t_0}^{t} d t^{\prime}
\gamma_{\beta}^{\K}(t^{\prime})
\left[p_{\beta}^{\alpha_0,t_0}(t^{\prime})\right]^2
e^{-\int_{t^{\prime}}^{t} d t^{\prime \prime}
\gamma^{\K}(t^{\prime\prime})},
\end{eqnarray}
respectively.
Here, $\gamma^{\K}(t)=\gamma_{+1}^{\K}(t)+\gamma_{-1}^{\K}(t)$,
$\Delta \gamma^{\K}(t)=\gamma_{+1}^{\K}(t)-\gamma_{-1}^{\K}(t)$, and
$p_{\beta}^{\alpha_0,t_0}(t)$ is obtained by solving the master
equation
\begin{equation}
\label{masterequation2}
\dot{p}_{\beta}^{\alpha_0,t_0}(t)=-\gamma_{\beta}^{\K}(t)
p_{\beta}^{\alpha_0,t_0}(t)+\gamma_{-\beta}^{\K}(t)
p_{-\beta}^{\alpha_0,t_0}(t),
\end{equation}
with initial condition $p_{\beta}^{\alpha_0,t_0}(t_0)=\delta_{\alpha_0,\beta}$.

In order to obtain expressions independent of the initial preparation,
 it is necessary to take the limit $t_0\rightarrow -\infty$ of
 Eqs.~(\ref{freqoutres1}) and (\ref{phasedis9}). In this limit, it can
 be shown that the functions
 $\Omega_\mathrm{out}(t)=\lim_{t_0\rightarrow -\infty}
 \Omega_\mathrm{out}^{\alpha_0,t_0}(t)$ and $
 D_\mathrm{out}(t)=\lim_{t_0\rightarrow -\infty}
 D_\mathrm{out}^{\alpha_0,t_0}(t)$ are periodic functions of the time
 $t$. Then, one can perform a cycle average and define the averaged
 output frequency
\begin{equation}
\label{finaloutfreq}
\Omega_\mathrm{out}=\frac{1}{T}\int_{0}^{T}
dt\,\Omega_\mathrm{out}(t)=\frac{\pi}{T} \int_{0}^{T} dt
\,\left[\gamma_{+1}^{\K}(t) p_{+1}(t)+\gamma_{-1}^{\K}(t)
p_{-1}(t)\right],
\end{equation}
where $p_{\beta}(t)$ is the periodic long-time solution,
$p_{\beta}(t)=\lim_{t_0\rightarrow -\infty}
p_{\beta}^{\alpha_0,t_0}(t)$, of Eq.~\eqref{masterequation2}. After
some lengthy calculations, it is also possible to show from
Eq.~\eqref{phasedis9} that the averaged phase diffusion is given by
\begin{eqnarray}
\label{phasediff2}
D_\mathrm{out}&=&\frac{1}{T}\int_{0}^{T} dt\, D_\mathrm{out}(t)=\pi\,
\Omega_\mathrm{out}
-\frac{\pi^2}{T}\mathrm{csch}\left(\frac{\bar{\gamma}^{\K}
  T}{2}\right)\nonumber\\&&\times \sum_{\beta=\pm 1} \beta
\int_{0}^{T}d t\int_{0}^{T} d t^{\prime} \Delta\gamma^{\K}(t)
\gamma_{\beta}^{\K}(t^{\prime})\left[p_{\beta}(t^{\prime})\right]^2
e^{\mathrm{sgn} (t-t^{\prime})\frac{{\bar{\gamma}^{\K}}
    T}{2}-\int_{t^{\prime}}^{t} d t^{\prime \prime}
  \gamma^{\K}(t^{\prime\prime})},
\end{eqnarray}
with
\begin{equation}
\label{averagedrate}
{\bar \gamma}^{\K}=\frac{1}{T} \int_{0}^{T} dt \,\gamma^{\K}(t).
\end{equation}
In the next section we will consider the case of a rectagular input
signal. In this case, explicit analytical evaluations of the integrals
in Eqs.~\eqref{finaloutfreq} and \eqref{phasediff2} can be carried
out.

\section{Periodic rectangular input signal}
As an example of the use of Eqs.~(\ref{finaloutfreq}) and
(\ref{phasediff2}) which is amenable to analytical treatment, we will
consider the case of the periodic rectangular driving force
\begin{equation}
\label{RF}
F(t)=(-1)^{n(t)} A,
\end{equation}
where $n(t)=\lfloor 2\, t/T \rfloor$, $\lfloor z \rfloor$ being the
floor function of $z$, i.e., the greatest integer less than or equal to
$z$. In other words, $F(t)=A$ [$F(t)=-A$] if $t\in [n T/2, (n+1)T/2)$
with $n$ even (odd). Because the potential fulfills the symmetry property
$U(x,t+T/2)=U(-x,t)$, we have $q_M(t)=(-1)^{n(t)}q_M(0)$, and
\begin{equation}
\label{minima}
q_\beta(t)=(-1)^j \frac{\Delta q(0)}{2}-(-1)^{n(t)}\frac{q_M(0)}{2},
\end{equation}
where $\Delta q(0)=q_{+1}(0)-q_{-1}(0)$. Here, we have taken
into account Vieta's formula $q_1(t)+q_2(t)+q_M(t)=0$.

According to the above mentioned symmetry property of the potential,
$\gamma_{\beta}^{\K}(t)$ can be expressed in the form
\begin{equation}
\gamma_\beta^{\K}(t)=\frac{\gamma}{2} \left [1-(-1)^{n(t)+\beta} \Delta
  p^{eq}(0) \right ],
\end{equation}
where $\gamma=\gamma_{+1}^{\K}(0)+\gamma_{-1}^{\K}(0)$, and $\Delta
p^{eq}(0)=p_{+1}^{eq}(0)-p_{-1}^{eq}(0)$, $p_{\beta}^{eq}(0)$ being
the equilibrium population of the state $\beta$ corresponding to the
rates taken at time $t=0$, i.e.,
$p_{\beta}^{eq}(0)=\left[\delta_{\beta,-1}\,
\gamma_{+1}^{\K}(0)+\delta_{\beta,+1}
\,\gamma_{-1}^{\K}(0)\right]/\gamma$. Notice that for the rectangular
input signal in Eq.~(\ref{RF})
$\gamma^{\K}(t)={\bar\gamma}^{\K}=\gamma$. We can also write
\begin{equation}
\Delta\gamma^{\K}(t)=\gamma_{+1}^{\K}(t)-\gamma_{-1}^{\K}(t)=-(-1)^{n(t)}
\gamma \Delta p^{eq}(0).
\end{equation}
As shown in Ref.~\cite{JCPPRL}, the long-time probabilities $p_{\pm
1}(t)$ are given by
\begin{eqnarray}
\label{persol}
p_{-1}(t)&=&\frac{1}{2} \left[1-(-1)^{n(t)} \Delta
p^{eq}(0)\right]\nonumber\\&& +(-1)^{n(t)} \Delta p^{eq}(0)\,
\frac{e^{-\gamma \left[t-\frac{n(t)T}{2}\right]}}{1+e^{-\frac{\gamma
T}{2}}},
\end{eqnarray}
and $p_{+1}(t)=1-p_{-1}(t)$.

Replacing the above expressions into Eqs.~(\ref{finaloutfreq}) and
(\ref{phasediff2}), one obtains after some lengthy simplifications that
\begin{equation}
\label{omegaout3}
\Omega_\mathrm{out} = \frac {\pi \gamma}2 \left \{ 1+ \left [ \Delta p^{eq}(0)
    \right ]^2 \left [ 1- \frac{4 \tanh \left (\frac{\gamma T}{4} \right
    )}{\gamma T}\right ]\right \},
\end{equation}
and
\begin{eqnarray}
\label{dif3}
D_\mathrm{out}&=& \pi \,\Omega_\mathrm{out} - \frac{2 \pi^2}{T}
\left[\Delta p^{eq}(0)\right]^4 \left[\tanh\left(\frac{\gamma
T}{4}\right)\right]^3 \nonumber \\ &&+ \frac{\pi^2}{2 T}\left[\Delta p^{eq}(0)\right]^2
\left\{1- \left[\Delta p^{eq}(0)\right]^2\right\} \nonumber\\
&&\times \left\{12\tanh\left(\frac{\gamma T}{4}\right)- \gamma T \left[ 1+2
\left[\sech\left(\frac{\gamma T}{4}\right)\right]^2\right]
\right\}.
\end{eqnarray}
In the next subsection we will compare these analytical results for the
averaged output frequency and phase diffusion with results obtained
from a numerical solution of the stochastic differential
equation~(\ref{SDE}).

\subsection{Comparison with numerical results}

Following the algorithm developed by Greenside and Helfand
\cite{Helfand,Greenside} (consult also the Appendix in
Ref.~\cite{JCP2003}), we have integrated Eq.~(\ref{SDE}) for a large
number of noise realizations, $M$, starting from one of the minima
$q_{\alpha_0}(0)$. From the initial instant of time, which we set
equal to zero, we start monitoring the switches of states and
recording the instants of time at which those switches occur,
according to Eq.~(\ref{def1}). We will denote by
$t_{n,i}^{\alpha_0,0}$ the instant of time of the $n$-th switch of
state in the $i$-th trajectory. From the switching times
$t_{n,i}^{\alpha_0,0}$, the realization of $N^{\alpha_0,0}(t)$
corresponding to the $i$-th trajectory, $N_{i}^{\alpha_0,0}(t)$, can
be easily calculated using Eq.~(\ref{def2}), and the corresponding
realization of the discrete phase, $\varphi_{i}^{\alpha_0,0}(t)$, by
Eq.~(\ref{defphase}). The noise averaged phase is then obtained by
\begin{equation}
\label{phaseaverage}
\left\langle \varphi^{\alpha_0,0}(t)\right\rangle =\frac{1}{M}
\sum_{i=1}^{M} \varphi_{i}^{\alpha_0,0}(t),
\end{equation}
and the phase variance by
\begin{eqnarray}
\label{phasedispersion}
v^{\alpha_0,0}(t)&=&\left\langle
\left[\varphi^{\alpha_0,0}(t)\right]^2\right\rangle -\left\langle
\varphi^{\alpha_0,0}(t)\right\rangle^2 \nonumber\\&=&\frac{1}{M}
\sum_{i=1}^{M} \left[\varphi_{i}^{\alpha_0,0}(t)\right]^2-\frac{1}{M^2}
\left[\sum_{i=1}^{M} \varphi_{i}^{\alpha_0,0}(t)\right]^2.
\end{eqnarray}
After a sufficiently long number of periods $L$ for the system to
``forget'' the initial preparation, the averaged output frequency is
calculated from the expression
\begin{equation}
\label{frequency2}
\Omega_\mathrm{out}=\frac{\left\langle
\varphi^{\alpha_0,0}\left[(L+1)T\right]\right\rangle-\left\langle
\varphi^{\alpha_0,0}(LT)\right\rangle}{T},
\end{equation}
and the averaged phase diffusion from
\begin{equation}
\label{diffusion2}
D_\mathrm{out}=\frac{v^{\alpha_0,0}\left[(L+1)T\right]
  -v^{\alpha_0,0}\left(LT\right)}{T}.
\end{equation}

\begin{figure}
\includegraphics[width=10cm]{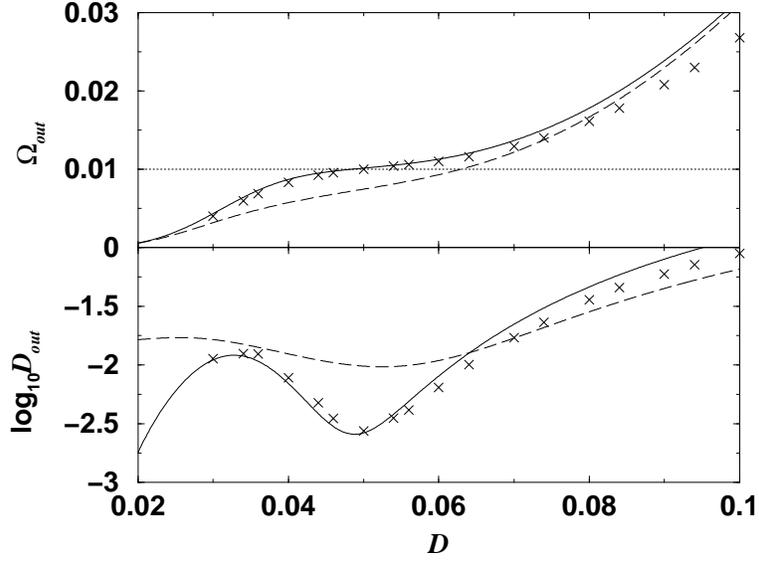}
\caption{\label{Fig2} Averaged output frequency (upper panel) and
  averaged phase diffusion (lower panel) as a function of the noise
  strength~$D$ for a periodic rectangular input signal [see
  Eq.~\eqref{RF} and text below] with amplitude $A=0.14$ and angular
  frequency $\Omega=2 \pi/T=0.01$ .  Solid line: Analytical results
  obtained from Eq.~\eqref{omegaout3} (upper panel) and
  Eq.~\eqref{dif3} (lower panel), respectively. Dashed line:
  Theoretical result from Ref.~\cite{Freund}. Crosses: Precise
  numerical results.  In the upper panel, a horizontal dotted line indicates
  the frequency of the input signal.}
\end{figure}

\begin{figure}
\includegraphics[width=10cm]{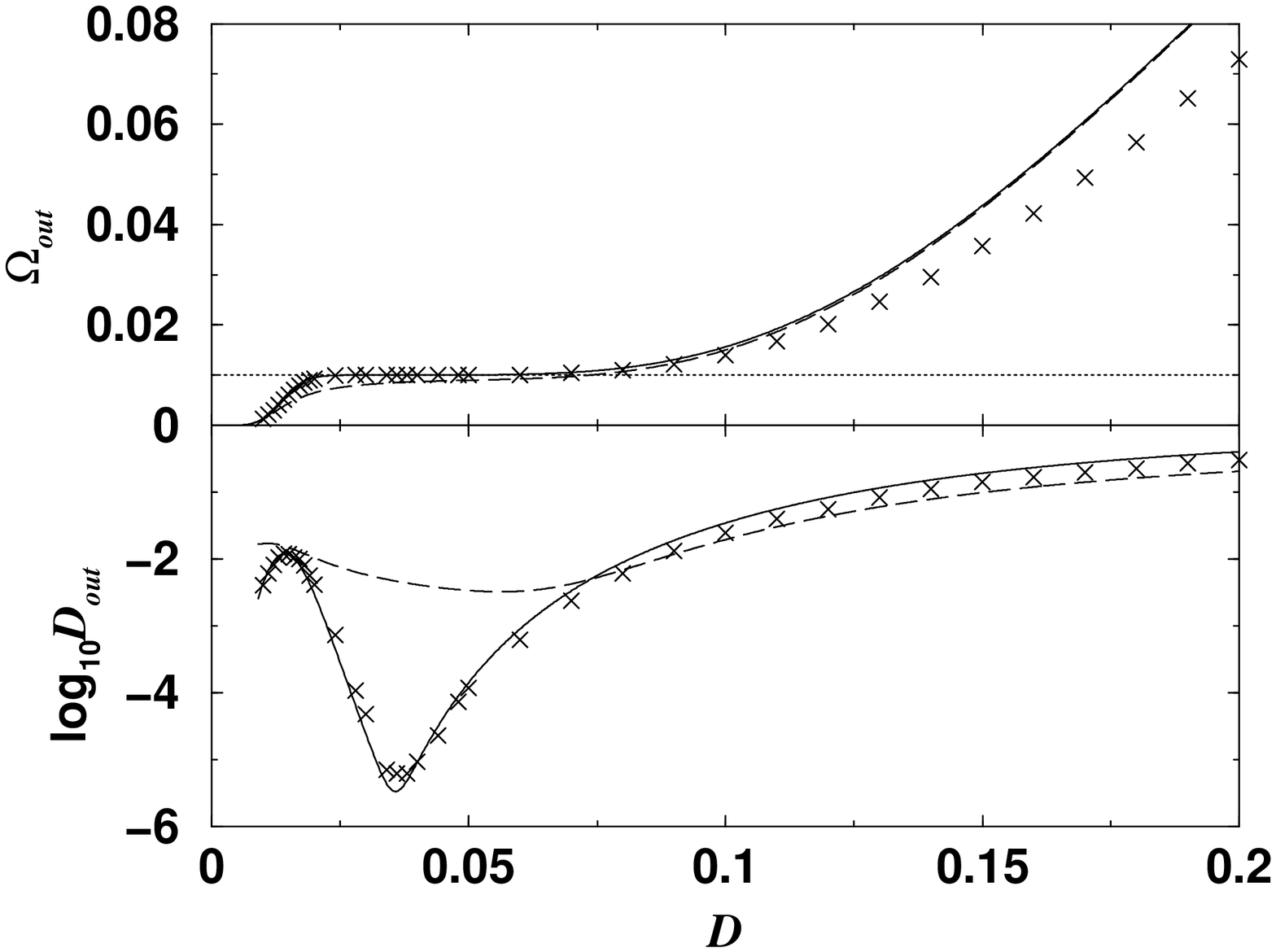}
\caption{\label{Fig3} Like Fig.~\ref{Fig2} but for a driving amplitude $A=0.25$.}
\end{figure}

\begin{figure}
\includegraphics[width=10cm]{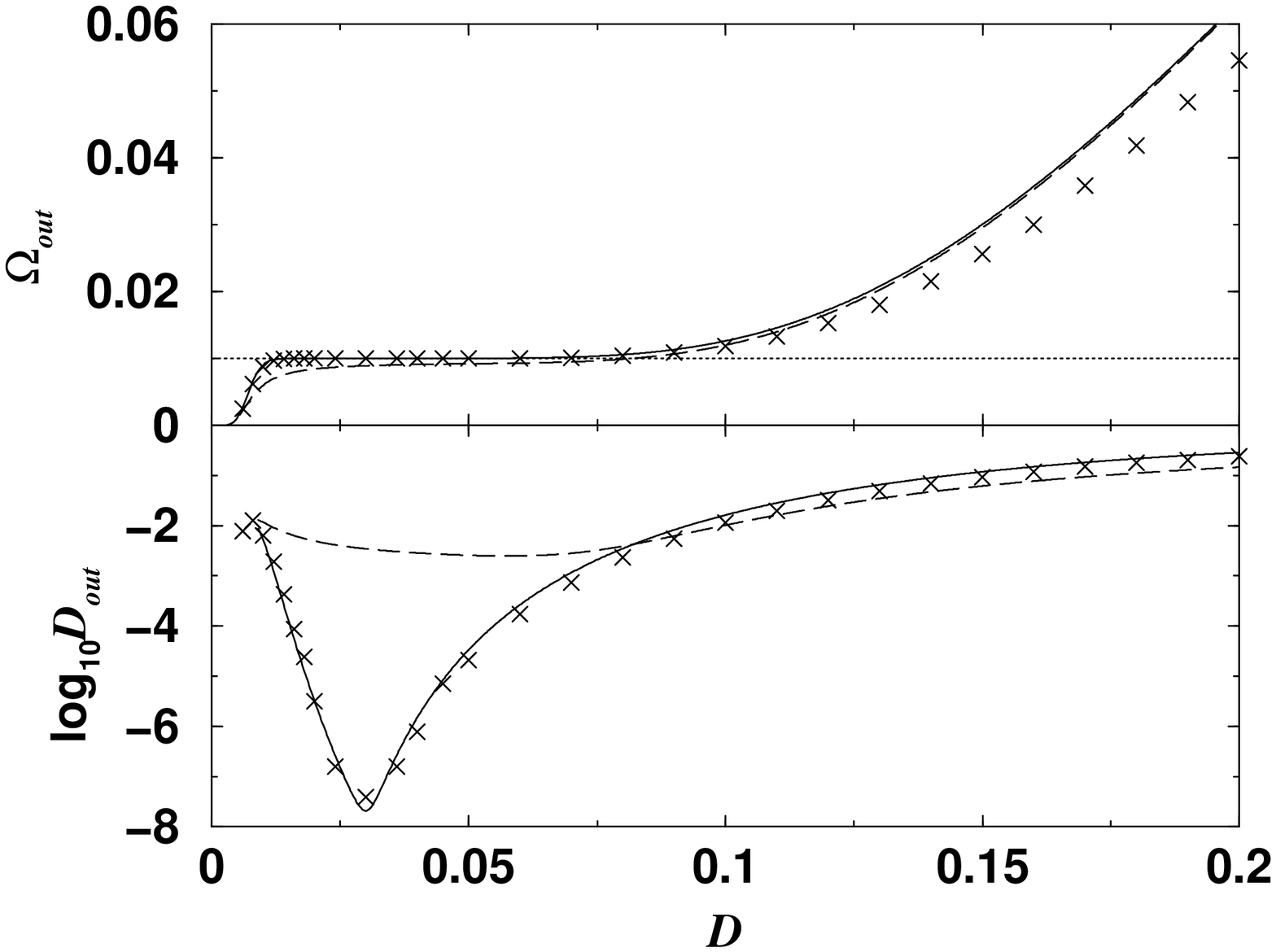}
\caption{\label{Fig4} Like Fig.~\ref{Fig2} but for a driving amplitude
  $A=0.30$.}
\end{figure}

Figures~\ref{Fig2}--\ref{Fig4} show the results of the numerical
solution just described for a periodic rectangular input signal with
angular frequency $\Omega=2 \pi/T=0.01$ and three different values of
the driving amplitude: $A=0.14$ (see Fig.~\ref{Fig2}), $A=0.25$ (see
Fig.~\ref{Fig3}), and $A=0.3$ (see Fig.~\ref{Fig4}).  As reported
previously in Ref.~\cite{Freund}, for high enough, but still
sub-threshold, driving amplitudes, the present system exhibits a
noise-induced frequency locking, i.e., starting from a non-zero value
of the noise-strength $D$, the frequency of the output signal matches
the frequency of the input signal, until, for strong noise, the output
signal becomes desynchronized again.  This effect is accompanied by a
very pronounced suppression of the phase-diffusion of the output
signal, i.e., a noise-induced phase locking. For the relevant values
of the noise strength $D$, our analytical estimates
Eq.~\eqref{omegaout3} and Eq.~\eqref{dif3} agree very well with the
results obtained from the numerical solution. Only for rather strong
noise, a noticeable deviation can be observed. In this regime, the
Kramers rates~\eqref{Kramers} are no longer valid. We have also
plotted the results of the previous work~\cite{Freund}, using the
rates~\eqref{Kramers}. We note that while the improvement of our
analytical estimates for the frequency synchronization is moderate
only, our prediction for the phase diffusion is strongly improved.


\section{Conclusions}

With this work we have investigated in detail the phenomenon of
frequency and phase synchronization in bistable, periodically driven
stochastic systems. This objective is not only of foremost interest
for the well known phenomena of Stochastic Resonance
\cite{gammaitoni1998,hanggi2002} and the topic of rocked Brownian
motors \cite{BM} but also carries great potential for the study of
driven stochastic neuronal dynamics and driven excitable sytems per se
\cite{lindner2004}. Our approach takes a new look on this prominent
problem. Starting out from a driven, Markovian continuous dynamics we
derived in great detail the stochastic renewal dynamics of the
noise-induced switching events.  This contraction of the full
Markovian dynamics in state space onto the discrete counting process
of subsequent switches between the metastable states implies a
non-Markovian dynamics for the switching times and the corresponding
phase dynamics whose explicit time-evolution depends on initial
preparation effects.  The resulting non-Markovian expressions still
contain the full information of the driven dynamics in the relevant state
space and thus are not readily accessible for analytical estimates.

In contrast, for weak noise and slow external driving, the dynamics of
the underlying process simplifies considerably.  Consequently, in the
long time limit the phase dynamics now assumes again a Markovian
nature. In this regime we put forward new results for the phase
diffusion and the frequency synchronization.  In doing so, we have
employed rectangular-shaped periodic driving signals.  This choice
entails two distinct advantages, namely (i) it allows a convenient
analytic analysis of the corresponding synchronization quantities and
(ii) its two-state character is also known to optimize the efficiency
for the synchronization features.  The same optimization feature holds
true for related effects such as the achievements of optimizing the
gain for Stochastic Resonance \cite{Casado2004} or the enhancements of
energy transduction in driven chemical reactions \cite{fulinski1992}.

Our analysis is in the spirit of prior works \cite{Freund,
  related-items}; our novel estimates, however, quantitatively
supersede in accuracy those prior results, cf. the detailed comparison
performed above. While the improvement for the frequency
synchronization is moderate only, the novel estimates present a
sizable improvement for the role of the phase diffusion. In the
weak-noise regime and for slow external driving, the Markovian theory
provides a very good agreement with numerical precise simulations.
This being so, we are confident that the new insight gained into the
complexity of non-Markovian, driven switching time dynamics together
with its Markovian simplification obtained at weak noise and slow
driving will prove useful for modelling and interpreting stochastic
synchronization phenomena in driven metastable and excitable dynamics.

\begin{acknowledgments}
  We acknowledge the support of the Direcci\'on General de Ense\~nanza
  Superior of Spain (BFM2002-03822), the Junta de Andaluc\'{\i}a, the
  DAAD program "Acciones Integradas" (P.H., M. M.), the
  Sonderforschungsbereich 486 (project A10) of the Deutsche
  Forschungsgemeinschaft and the Volkswagen-Stiftung (P.H.) under Grant
  No.\ I/77 217.
\end{acknowledgments}
%
%
% References
%
%

\end{document}